\documentclass[letterpaper, 10 pt, conference]{ieeeconf}  

\IEEEoverridecommandlockouts                              
\usepackage[utf8]{inputenc}
\usepackage[T1]{fontenc}
\usepackage{graphicx}
\usepackage{graphics}
\usepackage{xcolor}
\usepackage[normalem]{ulem} 
\usepackage[official]{eurosym}
\usepackage[switch,modulo]{lineno}
\usepackage[normalem]{ulem} 
\usepackage{footnote}

\makeatletter
\let\NAT@parse\undefined
\usepackage{hyperref}  

\title{\LARGE \bf
Regulation conform DLT-operable payment adapter based on trustless -- justified trust combined generalized state channels}

\author{Ricky Lamberty$^{1}$ and Alexander Poddey$^{2}$ 
\thanks{$^{1}$Center for Innovative Finance, University of Basel, Petersplatz 1, Basel, Switzerland
        {\tt\small ricky.lamberty@unibas.ch}}%
\thanks{$^{2}$Bosch Center for Artificial Intelligence, Robert Bosch GmbH, Corporate Research Campus, 71272 Renningen, Germany
        {\tt\small alexander.poddey@de.bosch.com}; 
        The project underlying this report was partially funded by the German Federal Ministry of Education and Research (BMBF) iBlockchain project under grant number 16KIS0904. The responsibility for the content of this publication lies with the author.
}%
}

\begin{document}

\maketitle
\thispagestyle{empty}
\pagestyle{plain}


\begin{abstract}

Open technologies, decentralized computation and intelligent applications enable the third-generation web, Web 3.0, thereby digitizing whole industries. The emerging Economy of Things (EoT) will be based on software agents running on peer-to-peer trustless networks that require a programmable, regulation conform means of payment. We give an overview of current solutions that differ in their fundamental values and technological possibilities, like e.g. private-issued stablecoins, DLT-issued electronic money and genuine cryptocurrencies. Based on this analysis, we present the concept of \textit{justified trust} and propose to combine the strengths of the crypto based, decentralized trustless elements with established and well regulated means of payment, based on this concept, via a secure external re-balancing interface.

Combining the advantages, e.g. lightweight, trustless, efficient high frequency micro state transfers on the one hand, and ease of use, widely spread, accepted alignment to a multitude of regulative requirements, on the other hand, while neither leading into a lock-in in any of the proposed solutions, nor undermining the basic principles of the crypto-movement or unnecessarily reinforcing the banking system provides a synergy and the necessary flexibility for further evolution alongside the regulative framework.
This offers a regulation conform transitional solution that can be implemented in the short term, which enables companies to place their decentralized business operations in a regulated environment. 

The contribution of our work is twofold: First, we illustrate and discuss different DLT-operable means of payment. Second, our research proposes a novel hybrid payment solution by interfacing trustless with justified trust combined generalized state channels.
\newline


\end{abstract}

\section{INTRODUCTION}
\label{sec:intro}

Due to improvements in crypto technologies adoption, it is now possible to convert the emerging Internet of Things (IoT) into an Economy of Things (EoT). Distributed Ledger Technology (DLT) is therefore capable to enable the EoT in a decentralized way. The Economy of Things refers to a heterogeneous hybrid digital economy of diverse participants e.g. IoT devices, digital entities, software agents running in the cloud and humans. This leads to the growth of more and more socio-economic aspects, since machines or digital representatives of humans will be able to enter binding agreements, and execute business transactions inclusive autonomous payments\cite{z11},\cite{z2}. 

A digitized economy cannot strive properly without an efficient settlement, payment functionality to enable the exchange of value. Today’s payment system are not built for either human-to-machine nor machine-to-machine transactions or immutable smart contract execution, while complying with all regulatory and legal requirements. Traditional financial service providers have struggled to evolve faster and cheaper payment services that can operate across borders. Even if the above-mentioned requirements could already be provided through technological adaptation of DLT, the nation-state specific regulative framework lags behind. Especially with respect to payment and the underlying value presentation through e.g. a cryptographic asset, a nation-state regulative needs to be involved to fight against money laundering and to sustain monetary sovereignty. However, for corporate contexts and widespread adaption there is still a quite high complexity and regulative issues\cite{z26}. For corporates, the handling and payment of cryptocurrencies causes expenses and poses additional risks. On the one hand, high costs for adaptation, e.g. issuing invoices in cryptocurrencies or adapting ERP systems. On the other hand, neither the customer nor the company is willing to accept the volatile exchange risk associated with cryptocurrencies. This outlines the necessity for a regulated, fiat-denominated payment solution that is processable on a decentralized network. The question is how to approach from ’where we are today’ with current standards and regulation in order to prepare a continuous evolution through transitional solutions.

Our paper provides an overview of current DLT-operable means of payment and highlights the tension between the different regulative, corporate's and trustless requirements when it comes to payment. In this context, we offer a solution combining established means of payment with new technological approaches, while both, being aligned with regulation and not leading into a vendor lock-in. The remainder of this article is structured as follows. In II, we highlight important characteristics of means of payments in a corporate context. Subsequent in Section III, we briefly describe general means of payment, whereas in Section IV we focus on DLT-operable means of payment. Based on our observation in Section IV, we propose a hybrid DLT-operable solution in V, and conclude the article in Section VI.

\section{The industry's payment needs}
\label{sec:industry}

Several requirements arise in order to get used as a means of payment for the emerging EoT. The following section provides an overview from a corporates perspective, since corporates need to fulfill a higher degree of regulation and verifiability. 
\subsection{Properties of a Currency}

A suitable currency should have following characteristics:
\begin{itemize}
    \item unit of account
    \item exchange medium
    \item and a store of value
\end{itemize}

Businesses and consumers don’t want to be exposed to unnecessary currency risk when transacting. Price stability thus provides a certain degree of planning certainty, whether with regard to production or currency risks. So far, cryptocurrencies have been suitable as an exchange medium, but not as a store of value. A token cannot represent a store of value if it’s price fluctuates by 20-30\% on a normal day\cite{z18}.
A company seeks to minimize risks, be it raw materials, investments or different means of payment. Especially in the financial sector, foreign currency positions represent a high risk potential. A company that uses cryptocurrencies as a means of payment aim to avoid additional risks, e.g. with regard to volatility. Certain instruments are used to minimize fiat or foreign currency risks, but this is always associated with additional effort.

\subsection{Legal and regulatory compliance}

An increasing number of companies are introducing cryptographic assets, or tokens, into their business that create new levels of risk. A lack of regulation can e.g. lead to new business models being compromised by late or incorrect regulation that carries additional business and reputational risks.
Europe has a set of regulation for dealing with cryptocurrencies, which is still in a state of change and therefor can change quickly from a regulatory point of view. The Financial Action Task Force recently issued cryptocurrency guidelines to ensure the technology is not used for financial crimes like money laundering and terrorist funding\footnote{In-depth: Guidance for a Risk-Based Approach to Virtual Assets and Virtual Asset Service Providers, https://bit.ly/3dMG4KY}.
In order to get widely accepted and used as a means of payment, a nation-wide regulation would be preferable. Detailed regulation will certainly be a long process, due to the fact that the technology itself is developing faster than the legislator’s regulatory attempt. This leads to an increase in complexity if it comes to cryptocurrency business adoption since no international standards exist\cite{z30}.

\subsection{Interoperable and Accessible}

Similar to the majority of financial innovations, which aim to reduce the frictions in the financial system, cryptocurrencies have emerged to address the existing market frictions stemming from the lack of a global trustless peer-to-peer payment mechanism. If the means of payment is limited to the use within one network, presented in \ref{sec:dltintermediaries}, multiple ecosystems will emerge. Within each of these ecosystems there is less friction as opposed to the friction between those ecosystems, comparable to the intranet and internet. This in turn means that companies have to connect their businesses to more ecosystems, in turn, this will create more friction. 
A successful means of payment needs a wide range of acceptance. User do not want to hold different e-Money token types, since no interoperability between those networks is currently given. Customers would need to hoard several means of payment in order to stay liquid, which in fact, is not economically viable. Therefore, interoperability is a main success driver of a means of payment. In addition, secure provisioning of the means of payment plays a crucial role in its acceptance, therefor minimizing the associated counter-party risk is inevitable. 

\subsection{Fiat-denominated (Accountable)}

Corporate denominate their invoices in fiat money, e.g. Euro in the EU, and therefore, also their accounting and ERP systems operate in Euro. The world at this point in time, in which "traditional" corporates issue an invoice in Bitcoin is not imaginable. Furthermore, a means of payment denominated in fiat currency does not pose additional effort to a company-specific accounting and clearing systems, since this systems already handle fiat-denominated transactions on a daily-basis.

\section{Means of Payment}

There are two types of means of payment. Legal tender and other means of payment. Banknotes issued by the European Central Bank and the national central banks are the only legal tender in Europe, everything else (Card payment, e-Money, Bitcoin), are other means of payment. In all countries, legal tender is subject to compulsory acceptance by the creditor (this is also referred to as an obligation to accept or a debt discharging obligation to accept). All other legally compliant means of payment are therefore free to choose between the contracting parties\footnote{§ 128 Abs. 1 Satz 3 AEUV}.

Money has established itself as the important means of payment which is used as a coordination device for facilitating transactions in human societies by fulfilling the role of a unit for the measurement of value across several goods and services. It is expected that the ongoing trend from direct human-to-human transaction to digital and even instant payments will further accelerate. In addition, due to the emerging digital human machine society, more and more demand for human-to-machine and even machine-to-machine payment will arise.
New forms of digital money and new means of payment are created, especially to address the need for cost-efficient instant- and micro-payments. The variety of digital means of payment has increased with the development of cryptocurrencies and is linked to a significant technological innovation. The continuous evolution of means of payments has lead to a multitude of possibilities ranging from physical nation-state to almost seamless non-nation state digital value transfer.

\section{Status Quo: DLT-operable Means of Payment}

The following section provides an overview of current \emph{DLT-operable} means of payment. \emph{DLT-operable} refers to the ability of being digitally processable on a DLT. The information needs to be transferred into a DLT network in order to allow automation through e.g. smart contract execution. Smart contracts would allow devices connected to a DLT, to provide services e.g. on a pay-per-use basis. The effects of a DLT-operable means of payment are therefore particularly promising in the context of the machine economy, as illustrated in section \ref{sec:intro}.

\subsection{Genuine Cryptocurrencies}

The lack of a global, censorship resistant currency allowing for cost-efficient international instant- and micro-payments has played a decisive role for the creation to some of the most prominent use-cases of blockchain, namely cryptocurrencies. Cryptocurrencies enable the peer-to-peer transfers of value in a trustless environment in a censorship-resistant manner\cite{z0}. 

Cryptocurrencies belong to virtual currencies, a clear distinction between regulated electronic money (e-Money) and virtual currencies can be found in the unit of account. E-money, which is bound by the traditional money format, has a grounded legal foundation as contrary to virtual currencies\footnote{The E-money directive (EMD) defines electronic money as “electronically, including magnetically, stored monetary value as represented by a claim on the issuer which is issued on receipt of funds for the purpose of making payment transactions…, and which is accepted by a natural or legal person other than the electronic money issuer”.}. This highlights the aspect that  dealing with virtual currencies, in contrast to e-Money, can lead to regulatory barriers.

Comparable to fiat currencies, and in contrast to e.g. gold, cryptocurrencies are created \emph{ex nihilo} and have no per se intrinsic value. However, fiat currencies are based on a nation state - or union of nations - specific issuance by a central authority, namely the central banks. Therefore, per construction, fiat currencies provide a centralized point of control. 

In contrast to that, cryptocurrencies can be understood as a digital representation of a market-side assigned value through the use of trustless technologies like DLT\cite{z1}. The construction, in a nutshell founding on non-nation-state related, decentralized control and transaction handling, is therefore fundamentally different. It is important to note that this difference actually mainly accounts for the market-side assigned value. 

The lack of centralized- and nation-state specific point of control, regarded as desireable properties by the community, however in turn is the reason for which cryptocurrencies are often considered a threat to the conventional, centralized banking system. The threat lies therefore in the unavailability of central, nation-state specific control, rather than being based on digital or even decentralized technologies itself. In fact there are examples of cryptocurrencies based on decentralized technologies, nevertheless providing a centralized point of control\footnote{e.g. Amazon Quantum Ledger Database (QLDB), https://aws.amazon.com/de/qldb/}. However, this undermines the basic principles of genuine cryptocurrencies and the main reason for their inherent value\cite{z20}.

Obviously, in case the adoption of decentralized controlled cryptocurrencies reaches a certain threshold, this could pose a threat to the established financial system, by having an impact on the realization of monetary policy\cite{z12} and transmission mechanisms\cite{z8}.

While the technological and fundamental, financial system related implications of such currencies, especially of Bitcoin, has attracted much attention, so far, there has been little discussion about the further limitations hindering a broad application in corporate contexts. Most prominently, volatility is still a common feature. The daily fluctuations, as well as the frequent spikes and crashes prevent these currencies among other reasons, to be used in daily transactions, such as buying goods or services. The usage of such currencies would mean an additional exchange risk, as mentioned in section \ref{sec:industry}.

As decentralized networks and cryptocurrencies become more widely adopted, corporations will need to ensure they can handle new payment methods and offer a variety of DLT-operable payment options. Companies need to set up a infrastructure to conduct a thorough legal and compliance analysis related to customer identity, taxation, liquidity, or other regulated issues. Additionally, accounting and cashflow systems need to be adjusted, especially if the company will be holding cryptocurrency for an extended time\cite{z19}. 

Nevertheless, Bitcoin and similar payment tokens have been classified as financial instruments by the Federal Financial Supervisory Authority (BaFin) in the form of units of account\footnote{§1 Abs.11 S.1 No. 7 KWG}, these tokens are classified equivalent to foreign currencies and can be used in today’s business transactions, but are not legal tender. This view is also shared by the European Banking Authority (EBA). 

\subsection{Stablecoins}

In this section we explain the basic properties of stablecoins. For an in-depth view we propose literature \cite{z4}, \cite{z16}, \cite{z15}. Stablecoins are a special representation of cryptocurrencies and solve the issue of volatility. This kind of cryptographic assets are developed with the aim of minimising price volatility by embedding a stability mechanism. Volatility is still a main hurdle for the widespread adoption as consumers and businesses require stability as a prerequisite for using any type of currency. In general, three different types of stablecoins exist:
\begin{itemize}
\item Collateralised stablecoins: On-chain
\item Collateralised stablecoins: Off-chain
\item Algorithmic stablecoins
\end{itemize}

The distinguishing characteristics between the three types are the stability mechanism and the nature of the collateral. Many stablecoins use the same mechanisms as current fiat currencies such as a currency basket, dollar pegs and a currency boards which are either centrally or decentrally managed\cite{z4}. A risk, agents who act as designated market-makers may have significant market power and the ability to bias the stablecoin price. This leaves open the potential for market manipulation. At the same time, there is an unsolved oracle issue, since the managed stablecoin relies on a third-party for price feeds in order to maintain the price of the stablecoin to an external peg\cite{z15}.

\emph{Global stablecoins} (GSCs), like Libra\footnote{In-depth: https://libra.org/}, can have far-reaching consequences from an economic point of view. GSCs widely used as a store of value, could have adverse effects on the transmission of monetary policy of domestic interest rates and credit conditions. Furthermore, implications of currency substitution would pose a threat to monetary sovereignty\cite{z16}. The Bank for International Settlement (BIS) has convened a G7 working group on stablecoins to better understand and assess the potential and risks. The G20 Financial Stability Board (FSB) has published a study on stablecoins, according to which they have the potential to be a threat to the global financial system. The German government, for example, has decided to oppose privately emitted stable tokens\footnote{In-depth: Federal Ministry of Finance, "Joint Statement on Libra", 2019}.This shows that regulators are even critical to non global constructions of stablecoins and cross-jurisdictional effort would be needed to combat money laundering and terrorist financing. 

In summary, it can be said that regulators struggle with the decentralized nature of genuine cryptocurrencies due to the lack of a central control element. Whereas with stablecoins, the threat lies in the steadily substitution of the domestic currency and the weakening of monetary transmission mechanisms. Given the regulatory landscape, it seems likely that regulators will not accept a stablecoin that has global scale\cite{z6}.

\subsection{DLT-issued e-Money}

The defining characteristic of e-Money is the digitized representation of traditional legal tender, but is no legal tender. From a traditional financial system's point of view, such an approach eliminates the downsides of genuine cryptocurrencies: the issuer provides a central point of control, is well aligned with the nation states monetary sovereignty, provides stability of value (with respect to the underlying fiat currency) and could be classified as legal tender. However, from a genuine trustless DLT-based point of view, e-Money is just that: a digitized representation of traditional, nation state legal tender associated with all deficiencies which the DLT movement seeks to overcome. It would be categorized a off-chain collateralised stablecoin. From such a point of view, the fact that DLT technologies are used for efficiency reasons does not, even not to some extend, address the main issues. 

We will call this type of currency, \emph{DLT-issued e-Money}, which can be regarded as nation-state specific, regulated, fiat-backed stablecoin or simply, programmable money. The following part explains the different design options on how "e-Money" can be issued on a DLT.

\subsubsection{DLT-issued e-Money by Intermediaries}
\label{sec:dltintermediaries}

Intermediaries are currently able to transform fiat currencies onto a DLT system by tokenizing fiat money deposited at banks or payment institutions. By intermediaries we mean regulated banks or e-money institutes, that are required to comply with strict safeguarding requirements to protect customers\footnote{In-depth: Article 7 of the E-Money Directive.} Comparable to the established banking system, the issuer provides a central point of control, and the central regulation forms the source of trust. Currently, there are a set of providers which e.g. issue Euro-denominated tokens on a private-permissioned networks. The reason for issuance on private-permissioned networks lies in the possibility for preserving central control over the underlying network and technology. Staying as close as possible to the established financal system's architecture, on the one hand is believed to be the most direct approach to legal compliance. However, the approach strongly deviates from the genuine cryptocurrency fundamentals.

To give two examples, the Commerzbank issues their e-Money token on R3 Corda\footnote{Main Incubator, https://bit.ly/3bp06Kj}, whereby the startup Cash-on-Ledger issues e-Money on the Alastria private-\footnote{Cash on Ledger, https://bit.ly/2YUksIF}. Both companies provide proprietary models of a programmable Euro that lack interoperability. Without a generally accepted standard, multiple fractional ecosystems will emerge and the advantage of money, being a unit of account and medium of exchange will vanish. Additional, this solutions could bear the risk of a lock-in effect for corporates.  

The DLT-issued e-Money must be 100 percent-backed, that means, reserves in fiat currency are stored off-chain in a bank account. Holders of these stablecoins must trust the issuer that all tokens are fully covered by fiat money deposits and that they could actually "pay out" their claim to fiat money in the same amount even in the case of an emergency. The central point of control offers a central point of attack since governments can simply shutdown by freezing the reserves held, if these issuer reach global scale. 
Furthermore, since the assets held as reserves are liquid, they pay no or very low interest, which offers less profitability for the issuer. In order to make money, the issuer of DLT-issued e-Money needs to engage in fractional reserve banking, which is then not much different from traditional banking\cite{z4}. 

\subsubsection{DLT-issued e-Money by Central Banks (CBDC)}

A CBDC, simply put, can be seen as digitised cash. The European Central Bank (ECB) defines CBDC as “a liability to a central bank that is made available to individual citizens in digital form”. CBDCs would thus represent a third form of central bank money, alongside banks’ reserves at the central bank and physical cash\cite{z14}. 

CBDC's would provide a central point of control and preserve nation states monetary sovereignty, on the one hand. Furthermore, the choice of design for a CBDC can even prevent the risks of the traditional fractional reserve banking system. By offering competition for bank deposits, the adoption of a CBDC could limit the practice of fractional reserve banking, thereby strengthening financial stability. To achieve this, various design approaches are currently being discussed. Therefore, with this regard they would, from a genuine DLT point of view, be preferable compared to intermediary issued e-Money. 
Regardless of the design, CBDCs would comply with current regulation\cite{z7}. Currency sovereignty is part of state sovereignty, so a central control point remains unavoidable. Whether or not the technology used is decentralized is irrelevant.

CBDCs might have broad implications  for the efficiency, stability and structure of the financial area. CBDC's would expand the variety for monetary policy mechanisms, e.g. variable interest rates would provide a new, non-redundant monetary policy instrument that could improve the overall effectiveness of monetary policy\cite{z13}. A central-issued CBDC could directly compete with commercial bank deposits, alleviate likely inducing a partial shift of deposits away from commercial banks towards the respective central bank. In turn, this might bear potential risks for the financial system.

This may significantly reduce the concentration of liquidity and credit risk in payment systems, resulting in a safer financial system, with less scope for impairment in monetary policy transmission\cite{z21}.

\section{Hybrid DLT-operable Payment Adapter}

\subsection{Preliminary Considerations}

From a corporate entities' point of view, as discussed in the foregoing sections, all the above means of payment are related to one or several drawbacks. To sum up, legal compliancy is a must and can best be achieved by staying as close as possible to the established banking system. However, fundamental values and  technological  possibilities related to DLT operable means of payment are quite tempting and form the basis for the transformation of the internet of things into a sound digital socio-economy.
 
For corporates, adopting a new form of payment in the day to day business is an involved and costly matter. Regulative issues, volatility and missing broad adoption by the normal user, on the one hand, and the risk to get locked-in a specific solution or having invested in the adoption of a non-prevailing technology, provide a context with much uncertainty, which in turn hinders adoption.

In order to overcome this, we propose to combine the strengths of the crypto based, decentralized trustless elements with established and well regulated means of payment. As we will discuss in the following, this is no contradiction leading to unnecessary overhead, but resembles a synergy combining the advantages (e.g. lightweight, trustless, efficient high frequency micro state transfers on the one hand, and ease of use, widely spread, accepted alignment to a multitude of regulative requirements, on the other hand) while neither leading into a lock-in in any of the proposed solutions, nor undermining the basic principles of the crypto-movement or unnecessarily reinforcing the banking system. In fact, the proposed approach can be interpret as a starting point providing simple adoption for private and corporate users, while allowing for seamless evolution. 

In the following section, we will outline the individual elements and discuss their fusion.

\subsection{Trustless Channels}
\label{sec:tresutlessChannels}

Naive application of DLT does not per se provide cost- and time efficiency. Using e.g. on-chain transfer of value via established blockchains is not in every case cost efficient, nor instant. Payment-, state- and generalized state channels are highly promising approaches, intended to reduce the number of required on chain interactions of distributed apps (dApps). A good overview of (generalized) state channels can be found in \cite{az1}\cite{az2}. The underlying interactions may be implemented in a trustless and secure manner, based on a number of different approaches like e.g. time- \& hash locks \cite{az6}\cite{az7}, forceMove\cite{az4} or Perun\cite{az3}. There also exist multiple extensions for increased on-chain efficiency and for N participants, see e.g. \cite{az5}\cite{z22}. 
All these constructions have in common, that they establish the possibilities for secure offchain (and therefore time- and cost efficient) transactions, while preserving security in a trustless way, by adequate anchoring on DLT systems. We refer to all of these constructions as \textit{trustless channels} in the following.

\subsection{Justified trust}
\label{sec:justifiedTrust}

The state of the art for trustless channels is, on the one hand quite powerful, however, requires all parties to actively take part in the channel and staking an adequate value (which is in fact the source of security).
This is not in every case desired by the participants of interaction networks (e.g. users dislike buying and holding volatile cryptovalues in order to be able to stake them for funding trustless channel constructions).

On the other hand, there exists a large set of solutions for \textit{secured} interaction in the pre-DLT world. The security thereby is not provided in a trustless/pre-funding fashion, but by regulation, laws, and contracts based upon this.

For example, bank regulation is involved (and often regarded as burden by the crypto community), however inter alia serves the purpose to provide trust for the client. Examples include deposit protection and the guarantee that the bank can not causelessly debit an amount of value from a clients account. 

Trust is provided by, from the viewpoint of the trustee-user relation, external enforcement of conformity with the given rules. In case one of the trustees (in the above example a bank) does not follow the rules, the customers, interest groups or responsible authorities can punish the misbehaviour and might even guarantee for a possible loss the customer might have otherwise. The user therefore, on a justifiably basis, can have trust in the trustee. 
Alternatively, the users might regard a certain level of trust as sufficient for certain interactions, e.g. transfer of a small amount of value, for wich the given level of trust is justifiable.

We refer to these settings in the following as \textit{\textbf{justified trust}}. More specifically, \textit{justified trust based on the regulative framework}, in case means for enforcement is provided by the regulative framework and \textit{justified trust due to reasonable context} in case the individual user regards the given level of trust reasonable in relation to the possible loss.

\subsection{Compliancy with regulations and established business practice}

In addition to the aspect of \textit{justified trust}, beneficial to the individual participants, the approaches for \textit{secured} interaction in the pre-DLT world are, in fact, involved solutions to a multitude of socio-economic requirements (in addition to bank regulation), like e.g. anti money laundering, and tax regulations. Although the crypto scene criticises the established solutions as too complex and rigid, the complexity and rigidness, at least a good part of it, is an effect of necessary fulfillment of these manifold regulative requirements.

On the other hand, for crypto based approaches, it is today at least unclear if and how the regulative requirements are to be applied. This is due to the fact that most often, the current form of the \textit{letter of the law} dates back to a pre-DLT time. Recent adaptions of regulations (like adaption of the European AML directive\footnote{Anti-money laundering (AMLD V) - Directive (EU) 2018/843}) however demonstrate that the \textit{spirit of the law} is just as well to be  applied to crypto based approaches. The \textit{letter of the law} will eventually reflect this, at the latest as soon as the business volume gets reasonably large.  

In addition, even though, smart contract tamper-proof execution provides the necessary verifiability in order to assert legal effects, and can be very useful in contractual settings, like automated execution of a contractual clauses, smart contracts are by definition no legal contracts and therefore not in every case sufficient for legal business. Smart contract based business models, operated by corporates therefore require to be embedded in a legal framework\cite{z28}.

\subsection{Interfacing trustless channels and justified trust systems}

\begin{figure}[h]
\includegraphics[width=8.6cm,height=10.5cm]{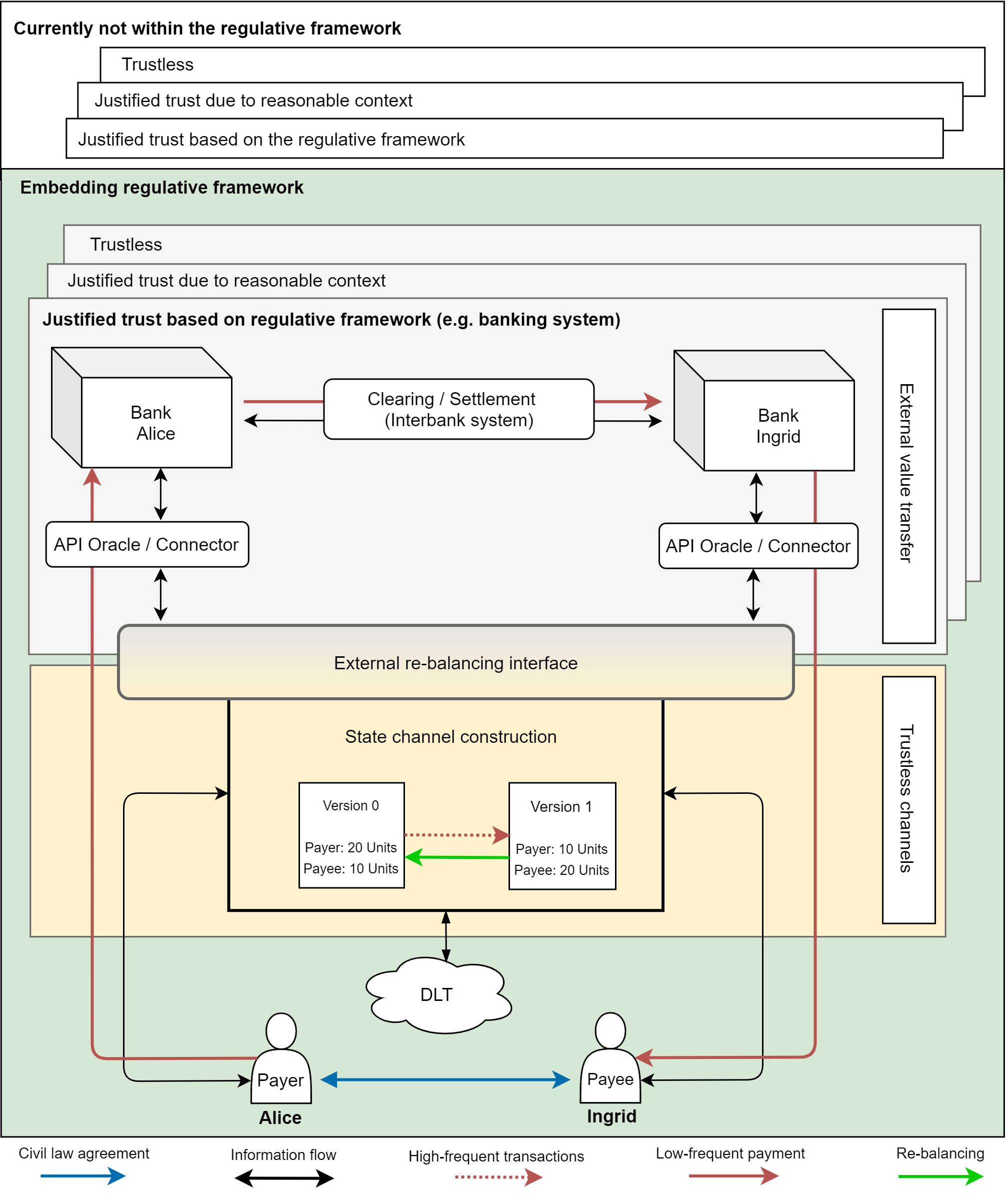}
\caption{\label{fig:arch}High-level architecture for trustless - justified trust combined systems. The connecting arrow coloured in blue illustrates a mutual agreement, based on the regulative framework. The arrows colored black indicate flow of information, whereas the elements coloured red consider the actual payment transaction. The green arrow represents the channel re-balancing. See the text for a discussion.}
\end{figure}

Based on these insights, we propose to interface trustless channels with justified trust systems as follows. Figure \ref{fig:arch} provides an illustration of the related architecture.

The regulative framework (illustrated as green area in Fig.\ref{fig:arch}) and legal contracts based upon it (blue arrow) provide an embedding context within which the on-boarding of the individual participants can be performed on a sound basis.

The high frequency interactions (dashdotted red arrow) between the individual participants are then handled efficiently via trustless channels (yellow area). 
The value assignment to the related crypto units of account, e.g. in relation to a given fiat currency, might, e.g. for straightforward conformance with existing regulations, be based on the underlying legal contracts. The application scope might also be well defined and possibly (initially) limited. This provides a steerable approach towards fulfillment of regulative requirements. In an extreme form, the trustless channel part can be regarded as crypto based, secure high frequency book keeping with no actual flow of value.

On a lower frequent basis (indicated by the red arrow), the resulting change of balance - related to an assigned value - can then be compensated via a channel-external, and in the context of the regulative framework established means of payment. We categorize the external means of payment in three basic classes (gray boxes), namely trustless, \textit{justified trust} based on the regulative framework and \textit{justified trust} due to reasonable context (see section \ref{sec:justifiedTrust} for details).

The interaction with the channel-external payment is designed such, that the corresponding channel is re-balanced accordingly (green arrow).
The required interface to the state channel thereby is realized by application of the usual building blocks of trustless constructions (like signed and counter-signed messages). Provided that the external message is based on another trustless system, or justified trust is given (e.g. a bank confirming receipt of an payment in favour of her client), the value transfer can be regarded as reasonable secure. Therefore, as usual with state channels, the individual user controls his security himself by accepting only external means of re-balancing which are based on reasonable trust. The required trust-level can be chosen on a per channel- or even per re-balancing basis.                                                                                                    

We provide detailed protocols for the external re-balancing interface in \ref{sec:appendixa}.

In a nutshell, secure interaction based on stake at risk and counter-signed and prooveable agreements, allows for dealing with dishonest participants and dispute resolution, as usual with trustless channels. These characteristics are effective up to the counterparts of the justified trust based payment system, e.g. the banks of Alice and Ingrid, respectively, in the example in Fig.\ref{fig:arch}. At that level however, the stake at risk for the counterparts of the external re-balancing (e.g. the banks), is not an explicit stake in the trustless channel, but is realized by the embedding regulative framework. In case of a dispute, every participant can proof the actual truth in a trustless fashion and thereby assert their rights, at any point in time. 
See \ref{sec:appendixa} for a detailed discussion. 

An example for an existing justified trust based transfer system would be the harmonized financial infrastructure of the eurosystem, discussed in section \ref{sec:appendixb}.	 

The steerability of regulation-related aspects, e.g. the application scope, and the required trust-level, as well as the flexibility to choose any (legal) form of payment the user agree upon, even on a per payment basis, allows to legally operate the solution concept on the basis of the current regulative framework, while providing the possibility for effortless continuous enhancement alongside the evolution of the regulation\footnote{In the example of Fig.\ref{fig:arch}, this would be related to inclusion of further instances of means of channel-external payment, currently non-existing or beeing in an unclear status relative to regulation, indicated by the white area on top, into the green area.}.  
In addition, the risk of getting locked-in a specific means of payment is strongly reduced.

\section{Conclusion and Outlook}

Fundamental values and  technological possibilities related to DLT operable means of payment are quite tempting and form the basis for the transformation of the internet of things into a sound digital socio-economy. However, for corporates, adopting a new form of payment in the day to day business is an involved and costly matter. Regulative issues, volatility and missing broad adoption by the normal user, on the one hand, and the risk to get locked-in a specific solution or having invested in the adoption of a non-prevailing technology, provide a context with much uncertainty, which in turn hinders adoption. Therefore, we argue, that, from a corpoate's point of view, the status quo of DLT-operable means of payment is related to several drawbacks.

In order to overcome this, we present the concept of \textit{justified trust} and propose to combine the strengths of the crypto based, decentralized trustless elements with established and well regulated means of payment, based on this concept, via a secure external re-balancing interface.

Combining the advantages, e.g. lightweight, trustless, efficient high frequency micro state transfers on the one hand, and ease of use, widely spread, accepted alignment to a multitude of regulative requirements, on the other hand, while neither leading into a lock-in in any of the proposed solutions, nor undermining the basic principles of the crypto-movement or unnecessarily reinforcing the banking system provides a synergy and the necessary flexibility for further evolution alongside the regulative framework. 

An already today practicable approach, based on our proposal, would e.g. be relating the value assignment to regulated, nation-state specific fiat currencies. Trustless channel technologies are then used to prevent fraud, while at the same time enabling high-frequency transactions. For the low-frequency, off-chain based re-balancing, regulators occupy the control points (Fiat-In, Fiat-Out), which provides compliancy to KYC/AML requirements. 
Such an approach would work well for use-cases where several micro transactions lead to a reasonable amount of value to be externally re-balanced on a low frequent basis, e.g. customer settlement on a weekly or monthly basis. Off-chain re-balancing would currently be problematic in case the amount is below minimal amount of value to be technically transferable (e.g. below 1ct in the banking system) or below the economic viable limit (due to the related cost of the transfer). Expressed pointedly, the approach is currently not best suited for singular micro transfer with the requirement of instant settlement. It is however to be noted that this limitation is due to regulation, and not caused by the underlying technology. On the contrary, the proposed approach provides high flexibility for efficient evolution alongside the regulative framework and preventing the emergence of multiple fractional ecosystems.  

Implementing operable solutions in compliancy with existing regulations already today, even in a quite limited form, in addition, allows to support the further development of the regulative framework in a constructive and consensual manner.

\section{APPENDIX}

\subsection{}
\label{sec:appendixa}

In the following, we illustrate the protocol steps for the external re-balancing interface based on a simple building block of trustless channels, namely a channel between Alice and Ingrid. We intentionally selected the name Ingrid as opposite participant to illustrate that our construction can be used as building block for more complex channel networks (in which Ingrid would e.g. act as intermediary). Our approach can be used together with a wide set of channel construction schemes. See section \ref{sec:tresutlessChannels} for references.

For simplicity, we start by assuming an already established (funded) trustless channel between Alice and Ingrid (established as usual in the individual protocols).
Note that the following extra steps for our extension could also be partly included in the opening / on-chain anchoring of the trustless channel and might be applied from simple channel constructions (no virtual channels or channel networks), up to n-participant, recursive virtual networks. 

We however illustrate the steps based on the current state of the art in which Alice and Ingrid are able to agree (without on-chain operations) on new rules, documented as smart contracts, vice versa signed-of by the participants. In addition, we only discuss a single state $X$ and the transfer from Alice to Ingrid (related to re-balancing $X$ from Ingrid to Alice).

Suppose Alice and Ingrid started with an initial state $X_A=20$, $X_I=10$ and had some interaction which resulted in $X_A=10$, $X_I=20$.

Usually, at some point in time the channel would be closed, allowing Ingrid to freely access and spend the received 10. 
However, in many circumstances, this is not desirable as:
\begin{itemize}
	\item it may be related to extra effort (e.g. on-chain transactions)
	\item it may be regarded as value transfer which is related to AML, tax and other regulations 
	\item the state value per se has no real value, but this value is only agreed upon by a (civil law based) contract between Alice and Ingrid (the trustless channel only provides a trustless, unforgeable way of bookkeeping for interactions between Alice and Ingrid)
\end{itemize}

To circumvent this, we make use of a justified trust based approach to let Alice send Ingrid the equivalent value of 10, in a channel-external way and securely re-balance the trustless channel accordingly (instead of closing it). To do so, care must be taken to prevent loss of value for any participant.

Suppose the justified trust based possibility (related as external transfer in the following) would be bank transfer. 
Alice is customer of Bank $B_A$, while Ingrid’s bank shall be $B_I$.
Suppose further Alice wants to transfer the equivalent value of 10 to Ingrid, such that the channel can be re-balanced to $X_A=20$, $X_I=10$.

The following steps need to be implemented in the underlying channel framework. The exact implementation depends on the channel construction, e.g. hash-time-locked, dApp based etc. The given description needs to be translated accordingly to the framework at use.

\subsection*{Initiation}
Alice informs Ingrid about her wish to achieve re-balancing of the channel via external transfer. Her message to Ingrid contains the channel specific information (depending on the specific channel construction) and, in addition, it may include a concretization about the method and conditions of the external transfer, the value equivalent etc.
Alice signs-of this proposal (as usual). We refer to this message signed by Alice as $m_{A1}$. 

If Ingrid agrees, she sends Alice a verifiable \textit{accepted} signal (e.g. a countersigned version of the original message from Alice) $m_{A1I1}$.

Once Ingrid agreed, $10+C_I$ of Ingrid are locked such that, after an agreed upon time, if nothing else happens, $10+C_I$ are credited back to $X_I$ (preventing loss or indefinitely long lock of her asset). $C$ thereby relates to an extra collateral, which could be any non negative value (including 0). Alice might analogously lock a collateral $C_A$.
The values are locked such, that, under certain circumstances (e.g. either providing or failing to provide a specific proof / certificate), a timer for timeout is reset or values are credited to the participants.

Note that, depending on the underlying channel construction the value might be transferred actively locked in a separate escrow contract, or only virtually (as with virtual channels \cite{az5}). 

\subsection*{Registration of valid sources of external information}
Assuming Alice and Ingrid agreed upon external value transfer via the banking system, $B_A$ and $B_I$ need to be registered in the channel as senders of certificates (e.g. via signed messages). This could either be anchored in the DLT level contracts, in any derived (virtual) channel, or in the concrete dApp anchored on the A-I channel, even on a per re-balancing procedure.

\subsection*{External transfer}
Next, Alice can trigger the external transfer to Ingrid.
Based on the level of presumed justified trust (e.g. Alice and Ingrid can trust the banking system as a whole, namely Ingrid can trust certificates issued by Alice’s Bank $B_A$ and vice versa or Alice trusts her bank and Ingrid $B_I$ only, etc.), several protocol alternatives for the external transfer are possible.

\textit{Alternative 1}

Alice and Ingrid have justified trust in the framework around $B_A$ and $B_I$.

Alice sends  $m_{A1I1}$ to $B_A$, $B_A$ checks if she is correctly registered as valid source of information. 
If not she can ignore Alices message (Alice could even get punished for false messages). If not further action by $B_A$ occur, after a timeout, Ingrid receives back her $10+C_I$, and Alice might loose her $C_A$ (transfered e.g. to Ingrid as compensation for the agreed transaction not beeing executed).

If however $B_A$ accepts $m_{A1I1}$, she triggers the transfer to $B_I$ (the security of this transfer is provided by justified trust in the banking system) and provides to Alice and Ingrid a certificate documenting that the transfer has been triggered ($Cert1$).

In a first design, this certificate could act as releasing the locked 10 to Alice and the collaterals back to the respective parties (as it is expected that the banking system securely transfers the equivalent value, and in case not, options for external enforcement are provided, see also the following discussion).

In a second design, this certificate triggers the transition to a second locking stage as follows.
We will have a look at the optimistic case first. If everything goes as expected, the value is transferred from $B_A$ to $B_I$ within $t_{actualTransfer}$. Once $B_I$ receives the value on behalf of Ingid, she confirms the receipt to Alice and Ingrid, e.g. provides a certificate $Cert2$\footnote{Note that $B_I$ is also registered as valid source of information.}. If the certificate is valid, the locked 10 are credited to Alice and the collaterals back to the respective parties.
In case $B_I$ and $I$ fail to confirm receipt of value within $t_{transferMax}$, while $Cert1$ is valid, Alice is credited $10+C_I$ (the extra collateral as compensation for the waiting). 

Due to regulation, $B_A$ and $B_I$ are part of a sub-system having rules and punishment for incorrect or missing transfer. For example in case $B_I$ does not credit the value to Ingrid’s acount, Ingrid can proof and enforce her claim by showing $Cert1$ to a regulative authority. Possible extra evidence from within the banking system can then be used to decide if e.g. $B_A$ did maliciously confirm but not execute the transaction, or $B_I$ did in fact receive but not credit the value to Ingid's account. The malicious party can then be punished.

\textit{Alternative 2}

Everybody trusts her bank only.

In this case, if $B_A$ accepts $m_{A1I1}$ as valid, she informs $B_I$ about the desired transfer via $m_{A1I1BA1}$. This message can also go to Alice and Ingrid and can reset the timer for the re-transfer of 10 to Ingrid (as Alice made a step, namely triggered the transfer).
If $B_I$ agrees, she documents this by sending $m_{A1I1BA1BI1}$.

This message could trigger the release of 10 to Alice and the collateral's back to the respective parties (comparable to the first design).

Alternatively, $m_{A1I1BA1BI1}$ could trigger the change of locking conditions, and a receive certificate (cert2 above) based unlocking could be applied, as described above.

\textit{Remarks}

There exist many different options for application of punishment via collateral, timing and release of fractions of locked values, depending on individual participant's behaviour, which can be choosen and agreed upon by the participants.

In addition, note that $B_A$ and $B_I$ might also take part in funding and locking in the underlying state channels. This however, on the one hand would require the external transfer partners to be part of the underlying legal contracting, which is not desirable from a viewpoint of flexibility. To recap section \ref{sec:tresutlessChannels}, the participants should be able to choose any legal external value transfer mechanism, possibly becoming available even after the channel network is established.
On the other hand, the banks already have stake at risk, and adjudication and punishment of misbehavior is provided by the regulative framework.

\subsection{}
\label{sec:appendixb}

The Eurosystem has built with the Single European Payment Area (SEPA) a harmonized financial infrastructure. The SEPA network would provide a justified trust system, that can be used as a source for an actual and external value transfer in a regulated manner, is available to all bank account holder, integrated into corporates infrastructure and "efficient". Modifications and extensions to the current operating SEPA schemes will evolve, based on their supplement rather than total replacement. SEPA money transfers can mainly be executed within one day, and sometimes almost real-time if we look at SEPA instant payment. In addition, in 2018, EBA Clearing set up a task force to develop a uniform European standard for Request-2-Pay (R2P) based on ISO 20022. By using Request-2-Pay (R2P), a trigger mechanism initiated by the payee, payment can be made more secure. This offers the advantage over the normal bank transfer that the payment cannot be reversed within 8 days. The payee thus has a lower risk of payment default. Such improvements thus enhance the efficiency and lower payment defaults within a justified trust system\cite{z27}.
\end{document}